\magnification=\magstep1

\font\eightrm=cmr8 
\font\eightsl=cmsl8
\font\ninebf=cmbx9

\def\dvol{{\,\rm d}^3}

\def\what#1{\widehat{#1}}

\newcount\chno \chno=1
\newcount\equno \equno=0
\newcount\refno \refno=0

\font\bigSIZE=cmbx12 at 14.4pt

\def\startbib{\def\biblio{\bigskip\medskip
  \noindent{\bf BIBLIOGRAPHY.}\bgroup\parindent=2em}}
\def\endbib{\edef\biblio{\biblio\egroup}}
\def\reflbl#1#2{\global\advance\refno by 1
\edef#1{\number\refno}
    \global\edef\biblio{\biblio\smallskip\item{[\number\refno]}#2\par}}

\def\eqlbl#1{\global\advance\equno by 1
  \global\edef#1{{\number\chno.\number\equno}}
  (\number\chno.\number\equno)}


\def\msimp#1#2{#1%
  \hbox to 0pt{\hskip 0pt minus 3fill
    \phantom{$#1$}\hbox to 0pt{\hss$#2$\hss}\phantom{$#1$}%
    \hskip 0pt minus 1fill}}


\font\tendouble=msbm10 \font\sevendouble=msbm7  
\font\fivedouble=msbm5

\newfam\dbfam
\textfont\dbfam=\tendouble \scriptfont\dbfam=\sevendouble
\scriptscriptfont\dbfam=\fivedouble

\mathchardef\dbB="7042 
\mathchardef\dbC="7043 
\mathchardef\dbD="7044 
\mathchardef\dbE="7045 
\mathchardef\dbF="7046 
\mathchardef\dbG="7047 
\mathchardef\dbH="7048 
\mathchardef\dbI="7049 
\mathchardef\dbJ="704A 
\mathchardef\dbK="704B 
\mathchardef\dbL="704C 
\mathchardef\dbM="704D 
\mathchardef\dbN="704E 
\mathchardef\dbO="704F 
\mathchardef\dbP="7050 
\mathchardef\dbQ="7051 
\mathchardef\dbR="7052 \def\RR{{\fam=\dbfam\dbR}}
\mathchardef\dbS="7053 \def\SS{{\fam=\dbfam\dbS}}
\mathchardef\dbT="7054 
\mathchardef\dbU="7055 
\mathchardef\dbV="7056 
\mathchardef\dbW="7057 
\mathchardef\dbX="7058 
\mathchardef\dbY="7059 
\mathchardef\dbZ="705A 


\startbib
\noindent
\reflbl\jeans{
J.H. Jeans, {\it The Stability of a Spherical Nebula,} 
	Phil. Trans. Roy. Soc. (London) {\bf 199}, pp. 1-53 (1902).}

\noindent
\reflbl\fridmanpolyachenkoBOOK{
A.M. Fridman and V.L. Polyachenko, {\it Physics of Gravitating
	Systems}, vol. I \& II, Springer, New York (1984).}

\noindent
\reflbl\lifshitz{
E.M. Lifshitz, {\it } J. Phys. USSR {\bf 10}, pp. 116ff. (1946).}

\noindent
\reflbl\boernerBOOK{
G. B\"orner, {\it The Early Universe,} Springer, New York (1992).}

\noindent
\reflbl\weinbergBOOK{
S. Weinberg, {\it Gravitation and Cosmology}, New York,
 Wiley (1972).}

\noindent
\reflbl\chandraBOOK{
S. Chandrasekhar, {\it Hydrodynamic and Hydromagnetic Stability,}
	Oxford Univ. Press (1961). }

\noindent
\reflbl\binneytremaineBOOK{
J. Binney and S. Tremaine, {\it Galactic Dynamics,} Princeton
	Univ. Press (1987).}

\noindent
\reflbl\chandra{
S. Chandrasekhar, {\it The gravitational instability of an infinite
	homogeneous medium when a Coriolis acceleration is acting,}
	in: Vistas in Astronomy {\bf I}, pp. 344-347 (1954).}

\endbib


\centerline{\bigSIZE MATHEMATICAL VINDICATIONS OF} 
\centerline{\bigSIZE THE ``JEANS SWINDLE''}

\bigskip
\centerline{Michael K.-H. Kiessling}
\centerline{\it Department of Mathematics, Rutgers University}
\centerline{\it 110 Frelinghuysen Rd., Piscataway, N.J. 08854} 

\bigskip\bigskip
\noindent{\bf ABSTRACT:} 
The original Jeans dispersion relation and instability criterion 
are derived by a mathematically well-defined limiting procedure. 
The 
procedure highlights Jeans' physical reasoning and vindicates 
the$\!$ (in)famous
``Jeans swindle.'' A second, independent procedure
is stated which yields the same result.
\bigskip\bigskip\bigskip
\centerline{October 08, 1999}

\vfill
\smallskip
{\hrule\medskip
\noindent
$\msimp{\copyright}{c}$ (1999) The author. 
Reproduction for non-commercial purposes of this pre-print, 
in its entirety and by any means, is permitted. 
}

\eject

\noindent
{\bf 1. INTRODUCTION}
\hsize=16.5truecm

	In 1902,  J.H. Jeans [\jeans] derived his celebrated 
dispersion relation 
$$
\omega^2 = |{\bf k}|^2 c_s^2 - 4\pi G\rho_0
\eqno\eqlbl\jeansDR
$$
governing the evolution of infinitesimal disturbances of a fictitious 
infinitely extended, homogeneous and isotropic, static fluid of mass 
density $\rho_0$ that is coupled to Newtonian gravity.
	According to (\jeansDR), an initial disturbance whose wavelength 
$2\pi/|{\bf k}|$ is much smaller than the Jeans length
$$
\lambda_J = \sqrt{ \pi c_s^2 \over  G\rho_0}
\eqno\eqlbl\jeansL
$$
behaves essentially like a classical sound wave of sound speed $c_s$, 
with self-gravity contributing only slight corrections, but, as Jeans noted
[\jeans], perturbations whose wavelengths surpass the Jeans length 
can grow exponentially in time. 
	The dispersion relation (\jeansDR) has a counterpart in 
the kinetic theory of encounter-less stellar dynamics; see, for
instance, the monograph by Fridman and Polyachenko [\fridmanpolyachenkoBOOK]. 
	A related yet conceptually somewhat different question is
the evolution of infinitesimal disturbances of an expanding homogeneous,
isotropic general relativistic Friedman-Lema\^{\i}tre cosmology.
	E. Lifshitz [\lifshitz] found that disturbances with wavelengths 
larger than the `dynamical' cosmological Jeans length
$$
\Lambda_J = \sqrt{ \pi c_s^2 \over  G(\rho + p)}\, ,
\eqno\eqlbl\jeansLcosm
$$
grow like a power law, as compared to the background 
evolution, cf. [\boernerBOOK,\weinbergBOOK]. 

	The Jeans instability is generally regarded as forming the 
basis of our understanding of gravitational condensation.
	In particular, the non-relativistic criterion is invoked in 
astrophysical theories of the formation of stars and some smaller 
stellar systems, the relativistic one in cosmological theories 
about the formation of galaxies.

	Interestingly, while the analysis for the evolution of 	
infinitesimal disturbances of a homogeneous and isotropic 
{\it dynamical relativistic} universe proceeds in an orderly 
manner (see in particular [\boernerBOOK]), Jeans' original analysis 
for the evolution of infinitesimal disturbances of a homogeneous and 
isotropic {\it static non-relativistic} universe, which is reproduced
in pertinent textbooks and monographs on the subject, e.g. 
[\boernerBOOK-\binneytremaineBOOK], enjoys a rather questionable 
reputation.
	The following quotation is taken from the excellent monograph 
by James Binney and Scott Tremaine [\binneytremaineBOOK,  p. 287ff.] 
(emphasis in the original): 
\medskip

\hsize=15truecm

\hangindent=1.5truecm
\hangafter=0
{
\eightrm
\noindent
``We construct our fictitious infinite homogeneous equilibrium 
by perpetrating what we shall call the {\ninebf Jeans swindle} after 
Sir James Jeans, who studied this problem in 1902 (Jeans 1929). 
Mathematically, the difficulty we must overcome is that if the 
density and pressure of the medium $\rho_0$, $p_0$ are constant, 
and the mean velocity ${\bf v}_0$ is zero, it follows from Euler's 
equation (5-8) that $\nabla \Phi_0= 0$. On the other hand, Poisson's
equation (5-9) requires that $\nabla^2 \Phi_0 = 4\pi G \rho_0$. These
two requirements are inconsistent unless $\rho_0 =0$. Physically,
there are no pressure gradients in a homogeneous medium to balance
gravitational attraction. A similar inconsistency arises in an
infinite homogeneous stellar system whose DF is independent of
position. We remove the inconsistency by the {\eightsl ad hoc} assumption
that Poisson's equation describes only the relation between the
perturbed density and the perturbed potential, while
the unperturbed potential is zero. This assumption constitutes the 
Jeans swindle; it is a swindle, of course, because in general there is
no formal justification for discarding the unperturbed gravitational
field.''}
\medskip

\hsize=16.5truecm

	Since modern cosmological applications do not any more rely 
on the original Jeans analysis but operate with the unproblematical 
relativistic dynamical analysis instead, and since a procedure 
for which ``in general there is no formal justification'' 
does not make much mathematical sense, one might want to conclude
that the original Jeans analysis is best avoided at all. 
	However, the importance of the stability analysis of an idealized
infinite homogeneous and isotropic static universe lies not in real
world physics applications but in the prospect of simplifying the 
analysis of certain physically relevant questions while retaining 
some essential features of a real system. 
	As such it has been taught, tongue in cheek of course, to 
generations of students and newcomers to the field. 
	But if we cannot backup the Jeans swindle by a 
formally correct analysis, then we face the dilemma of 
rendering a questionable service, surely to the novice in the field, 
but ultimately also to ourselves.

	As way out of this dilemma, Binney and Tremaine suggest that 
in special situations a formal justification of the ``Jeans swindle'' 
might exist. They list two examples [\binneytremaineBOOK,  p. 288]: 
\medskip

\hsize=15truecm

\hangindent=1.5truecm
\hangafter=0
{\eightrm
\noindent
``However, there are circumstances in which the swindle is
justified. For example, $\phantom{aaa}$
(i)... [if] ... the wavelength ... is much smaller than the scale over 
which the equilibrium density and pressure vary ... the Jeans swindle 
should be valid for the analysis of small-scale instabilities.  

\noindent
(ii) ... a uniformly rotating, homogeneous system ... 
can be in static equilibrium in the rotating frame and no Jeans
swindle is necessary (although the stability properties are somewhat
modified from those of the non-rotating medium because of Coriolis forces...)''
}
\medskip

\hsize=16.5truecm
	Unfortunately, upon closer inspection neither point (i) nor (ii) 
really justifies the Jeans swindle.
	Thus, (i) would work if the effective Jeans length of 
some self-gravitating equilibrium were much smaller than the scale 
of non-uniformity of that equilibrium. 
	However, the typical scale of non-uniformity  of
a self-gravitating equilibrium is precisely the effective Jeans
length, as emphasized in [\fridmanpolyachenkoBOOK], so that any
hypothetical ``small scale instability'' would have to be small 
in scale compared to the effective Jeans length, hence would 
{\it not} be analyzable by (\jeansDR). 
	In case of point (ii) the situation appears to be somewhat 
better, but also here there is a catch.  
	In fact, while the introduction of uniform rotation with angular 
frequency vector ${\bf \Omega}$ does regularize the homogeneous 
gravitational problem in such a way that an analysis in the 
spirit of Jeans can be carried out without invoking any 
`swindle' [\chandra], and while the resulting instability criterion 
for wave vectors satisfying ${\bf k}\cdot {\bf \Omega}\neq 0$ is 
precisely $|{\bf k}|^2 c_s^2 - 4\pi G\rho_0 <0$, in agreement with 
(\jeansDR), for this to justify the `Jeans swindle' we would now
have to be able to pass to the limit of a genuine `Jeans-swindle-situation' 
from this `no-Jeans-swindle-situation.' 
	But exactly that is not possible, because the angular frequency 
of a uniformly rotating equilibrium and the equilibrium mass density 
$\rho_0$ are related by $|{\bf\Omega}|^2 ={2\pi G\rho_0}$.
	As a consequence, the mass density $\rho_0$ has to vanish
in the non-rotating limit, and therefore a uniformly rotating system 
{\it does not} have a non-rotating limit in which the dispersion 
relation of the rotating system, which is different from (\jeansDR)
due to the presence of Coriolis forces [\chandra], would go over into 
the dispersion relation discovered by Jeans using his `swindle.'

	In this paper we will present a mathematically clean vindication 
of the genuine ``Jeans swindle,'' and a simple one at that, which emphasizes
the physics underlying Jeans' reasoning. 
	We also mention a second, independent method which gives the
same result.

	To begin with the physics, the first point to realize is that 
what counts dynamically are the forces, not the potentials. 
	Hence, as long as we obtain a sensible dynamics in 
some sensible limit, we should not worry too much if some 
potential ceases to exist in the same limit. 
	The second point to realize is that in Jeans'
originally conceived system, three infinities are
involved: (i) the infinite amount of matter, (ii) the 
infinite extend in space, and (iii) the infinite 
range of the Newtonian gravity. 
	Of course, (i) and (ii) are not independent because the assumption 
that the putative equilibrium is homogeneous, and the perturbations
only infinitesimal, couples these two infinities.
	However, (iii) is an independent source of trouble. 
	Therefore, we are well advised to attempt the construction
of the infinite system through a double limit.
	There are two mathematically natural possibilities:

(A) We first let the mass and size go  to infinity 
in a system with `screened' gravitational forces, and subsequently
`switch on Newtonian gravity' by removing the screening. 
	This treatment is borrowed from the class of infrared problems 
well known in quantum field theory.
	The standard procedure of handling infrared divergences 
is to apply an infrared regularization, to solve the regularized 
problem, and to remove the regularization at the end 
of the calculation, perhaps involving a `renormalization.' 
	As we will see in a moment, such a procedure works also here 
in an orderly manner.

(B) We study the system from the beginning with classical gravity, though
not in $\RR^3$ but in $\SS^3_R$, which is $\SS^3$ scaled to have
radius $R$. 
	Now the space is finite but without boundary, yet static. 
	We have no problem in defining a homogeneous and isotropic
static universe for classical gravity on $\SS^3_R$, and also not 
in studying the dynamics in its neighborhood. 
	Letting  $R\to\infty$ subsequently, using an obvious renormalization,
we  arrive at the Jeans dispersion relation for a system
in $\RR^3$, again in an orderly manner.

	In the rest of the paper, we will explain procedure (A)
explicitly, leaving (B) as (easy) exercise for the interested reader. 
	We first introduce the screened
gravitational interactions and explain that this does not lead
to physically unreasonable conclusions.
	Then we introduce the fluid-dynamical 
equations of an asymptotically homogeneous system with 
gravitational interactions by taking the no-screening limit
of the corresponding model with screened interactions. 
	The linearized version is precisely the system discussed by Jeans,
and his dispersion relation (\jeansDR) follows at once. 

	The equations of encounter-less stellar dynamics are
treated analogously as limit 
of an asymptotically homogeneous system with screened gravitational 
interactions, yielding the corresponding Jeans dispersion
relation for that model. 

\vfill\eject

\bigskip
\chno=2
\noindent
{\bf 2. SCREENED GRAVITATIONAL INTERACTIONS}
\medskip

	Consider first a sufficiently well behaved mass density function 
$\rho({\bf x})$ which, for the moment, shall have finite mass, thus
$\int_{\RR^3}\rho({\bf x}) \dvol x =M$.
	Replace the familiar Newtonian potential of $\rho$ at 
${\bf x}\in \RR^3$, given by
$$
\Phi({\bf x}) 
= 
-G\int_{\RR^3} {1\over |{\bf x - y}|} \rho({\bf y}) \dvol y\, ,
\eqno\eqlbl\newtonPOT
$$
by the screened Newtonian potential 
$$
\Psi({\bf x}) 
= 
-G\int_{\RR^3} {e^{-\kappa |{\bf x - y}|}\over |{\bf x - y}|} 
\rho({\bf y}) \dvol y
\eqno\eqlbl\yukawaPOT
$$
where $\kappa^{-1}$ is the screening length.
	In the limit of vanishing screening, i.e. $\kappa\to 0$, 
(\yukawaPOT) reduces to the Newtonian potential (\newtonPOT).
	Therefore, if $\kappa$ is tiny enough,  
for instance $\kappa^{-1} =\ size\ of\ visible\ universe$ say,
then for all concrete physical situations  where one can apply 
the Newtonian gravitational potential (e.g., 
non-relativistic planetary motion, stellar dynamics and even some 
galactic dynamics) one can as well apply the above screened 
gravitational potential.\footnote{$^\dagger$}{Of course, this 
  is not to say that a screened interaction with tiny 
  screening enjoys an equal physical status as the Newtonian 
  interaction -- quite the contrary is true. 
  By applying Occam's razor, there is no point in introducing
  screened interactions for the discussion of actual finite astrophysical 
  gravitating systems in the non-relativistic regime.}  

	Now drop the requirement that $\int \rho \dvol x =M$ and
consider a monotone sequence of mass densities 
that converges  to a constant mass density $\rho_0 >0$
(for instance, in sup norm, which means that
$\sup_{{\bf x }\in\RR^3}|\rho_0 - \rho({\bf x })|~\to~0$).
Then (\yukawaPOT) converges (likewise in sup norm)
to a constant limit as well, given by
$$
\Psi_0 = -G\rho_0 \int_{\RR^3} {e^{-\kappa |{\bf x - y}|}\over
|{\bf x - y}|} \dvol y = - 4\pi G\rho_0{1\over \kappa^{2}}
\eqno\eqlbl\nullPOT
$$
while the Newtonian potential diverges,\footnote{$^{\dagger\dagger}$}{
	In fact, $\Phi = -\infty$ even before $\rho$ has 
converged to $\rho_0$, which is a consequence of considering
convergence in sup norm and the fact that $|{\bf x}|^{-1}$ is
not integrable over $\RR^3$. 
	If instead of sup norm convergence one uses the weaker
concept of sup norm convergence on all compact sets (for instance
by considering a mass density which equals $\rho_0$ inside a 
ball of radius $R$ centered at ${\bf y}$ and which vanishes
outside the ball, and then letting $R\to \infty$) then once 
again  $\Phi\to -\infty$ though this time $\Phi$ stays finite 
for all finite $R$. Of course, $\Psi\to\Psi_0$ also now,
this time on compact sets.}
 $\Phi\to -\infty$, as
$\rho\to\rho_0$.
	The divergence of $\Phi$  as $\rho\to\rho_0$ is not yet bad 
news, though, for we know what counts are not the potentials but the 
forces derived from them, viz. their gradients. 
	Since a homogeneous system has no gradients, 
we would be in an acceptable limit situation if only we could guarantee 
that the gradient $\nabla\Phi$ would converge to zero as $\rho\to\rho_0$. 
	Unfortunately, not only does the gradient $\nabla\Phi$  {\it  not}
converge to zero, its limit (whenever it exists) does not just depend
on the limit density $\rho_0$ but on the particular limiting sequence 
$\rho\to\rho_0$ (a fact which is well known in astrophysics and
which has contributed to the general belief that 
Jeans' analysis is a `swindle.') 

	The existence of the constant limiting 
screened potential $\Psi_0$ on the other hand guarantees,
in conjunction with the definition (\yukawaPOT), that 
$\nabla\Psi\to\nabla\Psi_0 = {\bf 0}$ when $\rho\to\rho_0$.
	In other words, the screened gravitational interactions 
cancel themselves out when $\rho =\rho_0$. 
	Since no pressure gradients are needed to counterbalance  
these self-balanced screened gravitational forces, it follows that
such an infinite, homogeneous and isotropic fluid is automatically
in equilibrium. 
	The infinite homogeneous and isotropic equilibrium fluid with 
self-balanced gravitational forces can now be defined by simply 
taking the limit $\kappa\to 0$ of this equilibrium family. 
	
	In an analogous manner we can treat non-uniform mass density 
functions which differ from $\rho_0$ by the displacement of only a finite 
amount of mass. 
	We could be more general, but this is certainly a 
reasonable class of mass densities to study. 
	Hence, writing $\rho({\bf x}) = \rho_0 + \sigma({\bf x})$, 
the density disturbance $\sigma({\bf x})$ must be sufficiently
integrable, satisfy 
$$
\int_{\RR^3} \sigma({\bf x})\dvol x = 0\, ,
\eqno\eqlbl\neutrality
$$
and be bounded below by $-\rho_0$ (for $ \rho_0 + \sigma({\bf x})$
is a mass density and must therefore not be negative).
	For technical convenience, we actually demand that $\sigma$ 
be smooth and decay rapidly to zero at spatial infinity. 

	The screened gravitational potential $\Psi$ for such a 
mass density $\rho({\bf x}) = \rho_0 + \sigma({\bf x})$
is readily computed.
	By the linearity of the integral formula (\yukawaPOT), we have
$$
\Psi({\bf x}) = \Psi_0 + \psi({\bf x})
\eqno\eqlbl\yukawaSPLIT
$$
where $\Psi_0 = - 4\pi G\rho_0/ \kappa^{2}$ as before, and 
$$
\psi({\bf x})
= 
-G\int_{\RR^3} {e^{-\kappa |{\bf x - y}|}\over |{\bf x - y}|} 
\sigma({\bf y}) \dvol y
\eqno\eqlbl\disturbYUKAWA
$$
	Since $\sigma\neq\ constant$, the gradient of $\Psi$ in
general now does not vanish, but is given by
$$
\nabla \Psi({\bf x}) = \nabla \psi({\bf x})
\eqno\eqlbl\yukawaFORCE
$$
	Unless the force density $-\rho ({\bf x})\nabla \psi({\bf x})$ 
is counterbalanced by pressure gradients, $\rho({\bf x})$
will not be a stationary density.
	The ensuing time-evolution will be studied in the next 
section using the conventional Euler, respectively Vlasov equations
for such a system.

	The important point now is that because of the finite amount 
of mass involved in the density disturbance $\sigma$,  
the limit $\kappa\to 0$ of $\nabla\psi$ exists and is given by
$$
\lim_{\kappa\to 0} \nabla \psi({\bf x}) = \nabla \phi({\bf x})  
\eqno\eqlbl\newtonFORCE
$$
where
$$
\phi({\bf x})
= 
-G\int_{\RR^3} {1\over |{\bf x - y}|} 
\sigma({\bf y}) \dvol y
\eqno\eqlbl\disturbNEWTON
$$
is precisely the expression for the Newtonian potential of the
density disturbance invoked by Jeans. 
	Therefore our nonlinear dynamical equations for a
spatially {\it asymptotically} homogeneous and isotropic fluid 
(respectively, stellar system) with screened interactions turn 
-- without any swindle -- into nonlinear dynamical equations for 
an infinitely extended,  asymptotically homogeneous and isotropic 
fluid (stellar system) with Newtonian gravitational interactions. 
	Their linearization gives the equations originally 
derived -- in his heuristic manner -- by Jeans.

	We conclude this section with a few remarks regarding
Poisson's equation. 
	We notice that  the screened gravitational potential $\Psi$ 
given in (\yukawaPOT) satisfies the inhomogeneous Helmholtz equation  
$$
\Delta \Psi - \kappa^2 \Psi = 4\pi G \rho
\eqno\eqlbl\helmholtzEQ
$$
	Formally, as $\kappa\to 0$, the Helmholtz 
equation (\helmholtzEQ) for $\Psi$ goes over into the Poisson equation 
$$
\Delta \Phi =  4\pi G \rho\, 
\eqno\eqlbl\poissonEQ
$$
for the gravitational potential $\Phi$, {\it provided} the
solution $\Psi$ to Helmholtz's equation converges to a proper 
$\Phi$ when $\kappa\to 0$.
	Now, for the asymptotically homogeneous system where
$\rho = \rho_0 +\sigma$  and $\Psi = \Psi_0 +\psi$, we have 
$\Psi_0 \to -\infty$ when $\kappa\to 0$, as obvious from (\nullPOT). 
	Hence, since $\Psi \not\to\Phi$ in this case, 
Helmholtz' equation (\helmholtzEQ) for $\Psi$ does correspondingly
not turn into Poisson's equation (\poissonEQ) for some $\Phi$.
	However, once again this is not a problem because 
a constant potential is dynamically irrelevant. 
	Since we are entitled to `renormalize' the potential by 
subtracting an overall ($\kappa$-dependent) constant, we study 
$\Psi -\Psi_0$.  
	The Helmholtz equation for $\Psi -\Psi_0$ now does converge to 
a Poisson equation, namely the one which relates the density disturbance 
$\sigma({\bf x})$ to its gravitational potential disturbance $\phi$. 
	This is readily seen by recalling that
$\Psi({\bf x}) - \Psi_0 =  \psi({\bf x})$ is  the screened gravitational 
potential disturbance associated with the density disturbance 
$\rho({\bf x}) - \rho_0 = \sigma({\bf x})$, so that the 
Helmholtz equation for $\Psi -\Psi_0$ simply reads
$$
\Delta \psi - \kappa^2 \psi = 4\pi G \sigma\, .
\eqno\eqlbl\helmholtzEQdisturb
$$
	As $\kappa\to 0$, $\psi\to\phi$ , defined earlier
in (\disturbYUKAWA), and (\helmholtzEQdisturb) turns 
into the Poisson equation 
$$
\Delta \phi  = 4\pi G \sigma\, 
\eqno\eqlbl\poissonEQdisturb
$$
for the gravitational potential disturbance $\phi$. 
	All the {\it ad hoc} steps of the ``Jeans swindle'' have 
materialized in a mathematically clean way.

\vfill\eject
\bigskip
\chno=3
\noindent
{\bf 3. THE FLUID-DYNAMICAL JEANS DISPERSION RELATION}
\medskip

	After our discussion of the screened gravitational 
interactions, the derivation of the Jeans dispersion relation 
(\jeansDR) is now straightforward. 
	The model considered in this section is the Euler model 
of the inviscid motion of a fluid with screened self-gravitation. 
	The  variables of the model are the
fluid mass density $\rho$, pressure $p$, temperature $T$, 
fluid velocity ${\bf u}$, and the screened gravitational 
potential $\Psi$.
	The model equations comprise the continuity equation
$$
\partial_t \rho + \nabla \cdot (\rho {\bf u}) = 0\, ,
\eqno\eqlbl\contEQ
$$
the force balance equation
$$
 \rho\, \partial_t{\bf u} +  \rho {\bf u}\cdot \nabla\,  {\bf u} =
- \nabla p -\rho\nabla\Psi \, ,
\eqno\eqlbl\forceEQ
$$
an equation of state, for which (for simplicity) we choose
the law of the classical perfect gas at constant temperature $T_0$,
$$
p = {1\over m} \rho k_B T_0\, ,
\eqno\eqlbl\stateEQ
$$
and the Helmholtz equation for the screened gravitational potential $\Psi$,
$$
\Delta \Psi - \kappa^2 \Psi = 4\pi G \rho
\eqno\eqlbl\helmholtzEQb
$$
displayed here again for the sake of completeness.
	These equations have to be supplemented by
asymptotic conditions at spatial infinity, and initial conditions
at some initial time, say $t_0 =0$. 
	In particular, we demand that asymptotically at spatial 
infinity the dynamical variables approach the values of the 
stationary, infinite, homogeneous and isotropic equilibrium 
fluid in which the screened gravitational forces balance themselves.
	It is a trivial matter to verify that the set of constant variables, 
$\rho({\bf x})=\rho_0$, $p({\bf x}) = p_0 =(\rho_0/m)k_B T_0$, 
${\bf u}({\bf x}) = {\bf u}_0 = {\bf 0}$, and 
$\Psi({\bf x}) = \Psi_0 = - 4\pi G\rho_0/\kappa^2$
for all ${\bf x}$, forms such a stationary solution 
of (\contEQ-\helmholtzEQ).
	This constant solution will be our reference point.

	To inquire into the dynamics in the mathematical 
neighborhood of this constant solution, we
write $\rho({\bf x},t) = \rho_0 + \sigma({\bf x},t)$,
and demand that the initial $\sigma({\bf x},0)$ is smooth,
rapidly decaying to zero at spatial infinity, and satisfies
$$
\int_{\RR^3} \sigma({\bf x},0)\dvol x = 0\, .
\eqno\eqlbl\neutralityINIT
$$
	Then $\int_{\RR^3} \sigma({\bf x},t)\dvol x = 0$ for all 
$t\in (0,\tau)$, where $\tau$ is the mathematical lifespan of the
solution.
	Pressure and screened potential are represented accordingly, 
thus $p({\bf x},t) = p_0 + \sigma({\bf x},t)k_B T_0/m$, and 
$\Psi({\bf x},t) = \Psi_0 + \psi({\bf x},t)$. 
	We also write  ${\bf u}({\bf x},t)={\bf u}_0 + {\bf v}({\bf x},t)$.
	Although for our choice of reference equilibrium we have 
${\bf u}_0 =   {\bf 0}$, whence ${\bf u}({\bf x},t) = {\bf v}({\bf
x},t)$, we prefer to introduce a new symbol for the deviation from 
the equilibrium velocity field simply as a reminder that more
general equilibrium velocity fields can be handled. 

	Inserting the above representation of the dynamical variables	
into our fluid-dynamical equations, and already implementing
our equation of state into the force balance equation, as well
as using the fact that derivatives of constant functions vanish
and that $\Psi_0$ cancels versus $\rho_0$ from Helmholtz's equation, 
we obtain the dynamical equations for the unknowns $\sigma$, ${\bf v}$,
$\psi$, 
$$
\partial_t \sigma + \rho_0 \nabla \cdot {\bf v} 
+ \nabla \cdot (\sigma {\bf v}) = 0\, ,
\eqno\eqlbl\contEQdist
$$
$$
(\rho_0+\sigma)\, \partial_t{\bf v} + 
(\rho_0+\sigma)\, {\bf v}\cdot \nabla\,  {\bf v} 
=
- {k_B T_0\over m} \nabla \sigma 
-(\rho_0+\sigma)\nabla\psi
\eqno\eqlbl\forceEQdist
$$
$$
\Delta \psi - \kappa^2 \psi = 4\pi G \sigma\, .
\eqno\eqlbl\helmholtzEQdist
$$
All deviation variables are equipped with the asymptotic conditions
that they vanish asymptotically as $|{\bf x}|\to\infty$, all $t$.  

	At this point already we  can let $\kappa \to 0$ in
(\contEQdist-\helmholtzEQdist), thereby obtaining the nonlinear
dynamical equations for the evolution of the disturbances of an 
infinitely extended fluid with Newtonian gravity.
	The continuity equation remains unchanged, 
$$
\partial_t \sigma + \rho_0 \nabla \cdot {\bf v} 
+ \nabla \cdot (\sigma {\bf v}) = 0\, ,
\eqno\eqlbl\contEQdistNEWT
$$
while force balance and Helmholtz equation become
$$
(\rho_0+\sigma)\, \partial_t{\bf v} + 
(\rho_0+\sigma)\, {\bf v}\cdot \nabla\,  {\bf v} 
=
- {k_B T_0\over m} \nabla \sigma 
-(\rho_0+\sigma)\nabla\phi
\eqno\eqlbl\forceEQdistNEWT
$$
$$
\Delta \phi  = 4\pi G \sigma\, .
\eqno\eqlbl\poissonEQdist
$$
All deviation variables are equipped with the asymptotic conditions 
that they vanish at $|{\bf x}| \to\infty$.
	Notice that no linearization has been invoked so far.

	Proceeding on to the linearization of (\contEQdistNEWT-\poissonEQdist),
we write
$$
\eqalignno{
\sigma & = \sigma_1 + \sigma_2 + ...
&\eqlbl\sigmaEXPAND\cr
{\bf v}& = {\bf v}_1 + {\bf v}_2 + ...  
&\eqlbl\vEXPAND\cr
\phi &= \phi_1 +\phi_2 +...
&\eqlbl\phiEXPAND\cr}
$$
where the index $k = 1,2,3,...$ indicates the `level of smallness.'
	Thus, $\sigma_2$ is treated as one level smaller than $\sigma_1$; 
$\sigma_1\nabla\phi_1$ is at the same level of smallness as
$\rho_0\nabla\phi_2$; etc. 
	We are only interested in the first level of the hierarchy.
	Retaining only level 1 terms in (\contEQdistNEWT-\poissonEQdist),
we obtain the dynamical equations at level 1, 
$$
\partial_t \sigma_1 + \rho_0 \nabla \cdot {\bf v}_1 
 = 0\, ,
\eqno\eqlbl\contEQone
$$
$$
\rho_0\, \partial_t{\bf v}_1 
=
- {k_B T_0\over m} \nabla \sigma_1 
-\rho_0 \nabla\phi_1
\eqno\eqlbl\forceEQone
$$
$$
\Delta \phi_1  = 4\pi G \sigma_1\, .
\eqno\eqlbl\poissonEQone
$$
supplemented by initial conditions for $\sigma_1$, ${\bf v}_1$, and
the asymptotic condition of vanishing at infinity for 
$\sigma_1$, ${\bf v}_1$, and $\phi_1$. 
	The solution of these equations is carried out in the 
standard way using Fourier and Laplace  transforms, denoted by
$\widehat{\ }\ $ and $\widetilde{\ }\ $, respectively.
	For the density perturbation in particular we find
$$
\widetilde{\what{\sigma}}_1({\bf k},\omega) 
=
 {\omega \what{\sigma}_1({\bf k},0) 
- \rho_0 {\bf k}\cdot \what{\bf v}_1({\bf k},0) 
\over m^{-1}k_B T_0 |{\bf k}|^2 -4\pi G \rho_0 -\omega^2 }
\eqno\eqlbl\LFsigma
$$
from which we read off the original Jeans dispersion relation (\jeansDR)
for the disturbances of an infinite, homogeneous fluid with 
isothermal equation of state and Newtonian gravitational interactions.
	We have done so without invoking a `swindle,' or anything 
illegitimate of that sort.
\vfill\eject

\bigskip
\chno=4
\noindent
{\bf 4. THE STELLAR-DYNAMICAL JEANS DISPERSION RELATION}
\medskip

	In complete analogy, except that one has to be 
careful with the analytic continuations, one derives the stellar 
dynamical Jeans dispersion relation. 
	It suffices to summarize the main steps. 

	The dynamical variables of the model are, the particle density
function $f({\bf x},{\bf v},t)$ on the one-particle phase space 
$\RR^3\times\RR^3$ at time $t\in\RR$, and the screened gravitational 
potential $\Psi({\bf x},t)$.
	They satisfy Vlasov's dynamical equations, which comprise
the kinetic equation
$$
\partial_t f + {\bf v}\cdot\nabla f -\nabla\Psi\cdot\partial_{\bf v}f
=0
\eqno\eqlbl\vlasovEQ
$$
and the inhomogeneous Helmholtz equation for $\Psi$,
$$
\Delta \Psi - \kappa^2 \Psi = 4\pi G \int_{\RR^3} f\, {\rm d}^3 v
\eqno\eqlbl\helmholtzEQvlasov
$$
	Our reference solution consists of a phase space density that 
is homogeneous in physical space with mass density $\rho_0$ 
and Maxwellian in velocity space with temperature $T_0$, thus 
$f=f_0({\bf v}) =
 \rho_0 (2\pi k_B T_0/m)^{3/2}\exp(-m|{\bf v}|^2 /2k_B T_0)$,
and the Helmholtz potential 
$\Psi= \Psi_0 = - 4\pi G \rho_0/\kappa^2$, as before.
	Deviations from the stationary reference solution, written 
as $f({\bf x}, {\bf v},t) = f_0({\bf v}) + g({\bf x}, {\bf v},t) $
and $\Psi({\bf x},t) = \Psi_0 +\psi({\bf x},t)$, are required to
approach reference values at spatial and velocital infinity. 
	The dynamical equations for the  unknowns $g$ and $\psi$ are
$$
\partial_t g + {\bf v}\cdot\nabla g -\nabla \psi \cdot\partial_{\bf v}g
= \nabla \psi \cdot\partial_{\bf v}f_0
\eqno\eqlbl\vlasovEQdist
$$
$$
\Delta \psi - \kappa^2 \psi = 4\pi G \int_{\RR^3} g\, {\rm d}^3 v
\eqno\eqlbl\helmholtzEQvlasovdist
$$
	Taking the limit $\kappa\to 0$ gives the nonlinear Vlasov-Poisson
equations of an infinitely extended, asymptotically (in space) uniform
encounter-less stellar-dynamical system, 
$$
\partial_t g + {\bf v}\cdot\nabla g -\nabla \phi \cdot\partial_{\bf v}g
= \nabla \phi \cdot\partial_{\bf v}f_0 
\eqno\eqlbl\vlasovEQdistNEWT
$$
$$
\Delta \phi  = 4\pi G \int_{\RR^3} g\, {\rm d}^3 v
\eqno\eqlbl\poissonEQvlasovdist
$$
Expanding with respect to the levels of smallness,
$$
\eqalignno{
g & = g_1 + g_2 + ...
&\eqlbl\gEXPAND\cr
\phi &= \phi_1 +\phi_2 +...
&\eqlbl\phiEXPANDneu\cr}
$$
and retaining only level 1 terms, we find 
the linearized Vlasov-Poisson equations,
$$
\partial_t g_1 + {\bf v}\cdot\nabla g_1 
= \nabla \phi_1 \cdot\partial_{\bf v}f_0 
\eqno\eqlbl\vlasovEQlinear
$$
$$
\Delta \phi_1  = 4\pi G \int_{\RR^3} g_1\, {\rm d}^3 v
\eqno\eqlbl\poissonEQvlasovlinear
$$
whose solution in terms of Fourier and Laplace transformation is
standard. 
	In particular, the phase space density perturbation,
or rather its Fourier-Laplace transformed expression, 
for ${\cal I}m(\omega) <0$, reads
$$
\widetilde{\what{g}}({\bf k}, {\bf v}, \omega)
= 
{-i\,  \what{g} ({\bf k}, {\bf v},0)
	\over
\Bigl(\omega +{\bf k}\cdot {\bf v} \Bigr)
\biggl(\displaystyle{
1- {4\pi G m \rho_0 \over k_B T_0 |{\bf k}|^2} {1\over \sqrt{\pi}}
	\int_{-\infty}^\infty 
		{\xi e^{-\xi^2} 
			\over 
		\xi + {\omega \over |{\bf k}|}\sqrt{m\over 2k_B T_0 }}
{\rm d}\xi }\biggr)}
\eqno\eqlbl\vlasovLFdens
$$
which has to be analytically continued to ${\cal I}m(\omega) \geq 0$. 
	Apart from the ballistic term, absent in fluid theory, 
we immediately read off the stellar-dynamical Jeans dispersion relation 
for ${\cal I}m(\omega) <0$,
$$
1- {4\pi G m \rho_0 \over k_B T_0 |{\bf k}|^2} {1\over \sqrt{\pi}}
	\int_{-\infty}^\infty 
		{\xi e^{-\xi^2} 
			\over 
		\xi + {\omega \over |{\bf k}|}\sqrt{m\over 2k_B T_0 }}
{\rm d}\xi  =0
\eqno\eqlbl\jeansDRvlasov
$$
to be analytically continued to ${\cal I}m(\omega) \geq 0$. 
	In particular, using Plemilj's formula, we find that for 
${\cal I}m(\omega)=0$ the dispersion relation can be fulfilled only
if  ${\cal R}e(\omega) = 0$ as well, which gives as stability boundary
the familiar formula of Jeans,
$$
 |{\bf k}|_{_{J}}  =  \sqrt{{4\pi G m \rho_0 \over k_B T_0}}
\eqno\eqlbl\jeansKvlasov
$$
once again without any `swindle.'

\bigskip
\noindent
{\bf ACKNOWLEDGEMENT:} 
I thank my colleagues Carlo Lancellotti, Joel Lebowitz, and Sheldon
Goldstein for a careful reading of the manuscript. 



\biblio

\bye